\documentclass[12pt]{article}
\usepackage{amsmath}
\usepackage{amstext}
\usepackage{amssymb}
\usepackage{epsfig}
\usepackage{float}
\usepackage[square,sort,sort&compress,comma,nonamebreak]{natbib}

\usepackage[unicode,bookmarks,bookmarksopen,bookmarksopenlevel=0,colorlinks,linkcolor=blue,citecolor=red,hypertex]{hyperref}

\begin{document}
\thispagestyle{empty}
\begin{center}
\bf  Modeling of zonal electrophoresis in plane channel of complex shape
\end{center}

\begin{center}
\textbf{Shiryaeva E.\,V.}

Department of Mathematics, Mechanics and Computer Science, \\
Southern Federal University, 344090, Rostov-on-Don, Russia

shir@ns.math.rsu.ru

\bigskip
\textbf{Vladimirov V.\,A. }

Department of Mathematics, York University, York, YO10 5DD, UK

vv500@york.ac.uk

\bigskip
\textbf{Zhukov M.\,Yu.}

Department of Mathematics, Mechanics and Computer Science,
\\ Southern Federal University, 344090, Rostov-on-Don, Russia

zhuk@ns.math.rsu.ru
\end{center}

\section*{Abstract}\label{s:00}

\small The zonal electrophoresis in the channels of complex forms is considered mathematically with the use of
computations. We show that for plane S-type rectangular channels stagnation regions can appear that cause the strong
variations of the spatial distribution of an admixture. Besides, the shape of an admixture zone is strongly influenced
by the effects of electromigration and by a convective mixing. Taking into account the zone spreading caused by
electromigration, the influence of vertex points of cannel walls, and convection would explain the results of
electrophoretic experiments, which are difficult to understand otherwise.

\normalsize

\textbf{Keywords:} Microchip, electrophoresis.

\section*{Introduction}\label{s:0}

During the last ten years or so the intense use of various microchips aimed to separate mixtures by an externally
imposed electric field, to control micromixing and chemical reactions has been flourished (see
\emph{e.g.}~\cite{BerliPiaggioDeiber,BharadwajSantiagoMohammadi,ChenSantiago1, EricksonLi1,
EricksonLiuKrullLi,Ghosal47,GuijtEvenhuisMackaHaddad,Herr,HuWernerLi, JenWuLinWu,JohnsonRossLocascio,
KanianskyMasarBodor,MolhoHerrMosier,Santiago4,OddyMikkelsenSantiago, PatankarHu,ErmakovJacobsonRamsey2000,
ErmakovNano,ErmakovNano1,PatankarSantiago}). The industrial use of microchannels for the efficient separation of mixtures is well known
as the technology called Lab-on-a-Chip. The most effective control of mass transfer in microfabricated fluid devices
can be achieved with the use of electrokinetic phenomena such as electrophoresis and electroosmosis. The crucially
important part of related research is computer modeling that helps to improve the design of microchips, to understand
the processes involved, and to enhance experimental methods.

The modeling of the electrophoretic separation of a mixture represents a challenging problem due to the large number of
physical phenomena involved into the mass transfer driven by an electric field. One can count here such phenomena as
diffusion, chemical reactions, dependence of electrical conductivity on concentrations, electroosmosis, Joule heat,
convection,\emph{ etc.} It should be noticed that many papers devoted to the transport phenomena in microchannels take
into account only diffusion, electroosmosis, and the Taylor-Aris dispersion. As the result  these papers leave out of
account some essential nonlinear effects that appear due to the dependence of electrical conductivity on component
concentrations; it is well known that these effects significantly change zone shapes
\cite{BabZhukYudE,BelloZhRChrom95,ZhEREph96,ZhYuDAN,Zhukov2005,ZhJVM84}
and even can trigger a so called substance-lock effect \cite{ZhERSIAM} (for an experimental verification see
\cite{ZhECRAnal94}).

The effects of electric field singularities that occur near the vertex points of electrophoretic chamber walls still
have not been understood and mathematically described. For a simple case of a plane cross-shaped channel this
singularity is of the order $\mathcal{O}(r^{-1/3})$, where $r$ is a distance from the vertex point of the reflex angle
$3\pi/2$. The related distortions of zone shapes are described in \cite{BZhMyshkis} and experimentally
justified in \cite{ZhBSazStoyanov}. In addition, transport processes can be effected by convective mixing,
which can drastically deform the final stage of a separation process. The role of convection in electrophoresis is
described in
\cite{Bello,Polezaev-3,ZhZiva94,ZhSazonovDifUr97,ZhZiva95,ZhPetr97MZhG}. It is apparent that one can weaken
convection by the choosing of an appropriate orientation of a microchip in the gravity field. Nevertheless, taking
convection into account can not be avoided for high precision experiments.

This paper is devoted to the computer modeling that reveals the effects of such key factors as vertex points of channel
walls, electromigration, and convection on zone distortions. We present only the results of a small part of our
numerical experiments that have been carried out with the use of a specially created interactive program for the
modeling of zonal electrophoresis. The main result is the identifying of the parameter intervals when the distortions
of moving zones are the most significant. These data can be very useful for the planning of new experiments and for the
designing of electrophoretic chambers.

\section{Mathematical Model}\label{s:1}

The mathematical models of electrophoresis  (and in particular zonal electrophoresis) are well known
\cite{BZhMyshkis,BabZhukYudE,Zhukov2005,MosherSavilleThorman}. The dimensionless governing
equations describing both the motion of separated (by the action of an electric field) substances and fluid convection
(in the Oberbeck-Boussinesq approximation) are:
\begin{eqnarray}
&& \frac{d\boldsymbol{v}}{dt}=-\nabla p+\mu \Delta
 \boldsymbol{v} -  \boldsymbol{k}
 \sum\limits_{k=1}^{n}\beta_k c_k, \quad
 \operatorname{div} \boldsymbol{v}=0,
  \label{shir:1.1}
\end{eqnarray}
\begin{eqnarray}
&& \frac{d c_k}{dt}+\operatorname{div}
\boldsymbol{i}_k=0, \quad
\boldsymbol{i}_k=-\varepsilon|\gamma_k| \nabla c_k + \gamma_k c_k \boldsymbol{E},
  \label{shir:1.2}
\end{eqnarray}
\begin{eqnarray}
&& \boldsymbol{E}=-\nabla \varphi, \quad
 \boldsymbol{j}=\sigma\boldsymbol{E}, \quad \operatorname{div}
(\sigma \nabla\varphi)=0,
\quad (\operatorname{div} \boldsymbol{j}=0),
  \label{shir:1.3}
\end{eqnarray}
\begin{eqnarray}
&& \sigma=\sigma_0\left(1+\sum\limits_{k=1}^{n}\alpha_k c_k\right)>0,
  \label{shir:1.4}
\end{eqnarray}
Here $\boldsymbol{v}$ and $p$ are velocity and pressure,
      $\boldsymbol{E}$ and $\varphi$ are the strength and the potential of an electric field,
      $c_k$ is the $k$-th concentration ($k=1,\dots,n$),
      $\boldsymbol{i}_k$~--- the flux of concentration,
      $\boldsymbol{j}$~--- the  density of electric current,
      $\sigma$~---  mixture conductivity,
      $\sigma_0$~--- the conductivity in the absence of admixtures (the conductivity of a buffer solution),
      $\mu$~---  fluid viscosity,
      $\gamma_k$~---  electrophoretic mobility,
      $\beta_k$~--- the coefficient that appears in the linear dependence of density on concentration,
      $\alpha_k$~--- the coefficient that appears in the linear dependence of conductivity on concentration,
      $\varepsilon$~--- characteristic diffusion coefficient ($\varepsilon|\gamma_k|$ are diffusion coefficients),
      $ \boldsymbol{k}$~--- the unit vector of the $z$-axis that is anti-parallel to the gravity.
The dimensionless variables used are described in \cite{BZhMyshkis,BabZhukYudE}.

In the presented model the effects of Joule heat and electroosmosis have been neglected. The  conductivity of a mixture
has been modeled by the expression (\ref{shir:1.4}) that is natural for zonal electrophoresis; it is accepted that a
mixture contains components with constant concentrations that represent so called buffer solution (such that at $c_k=0$
its conductivity is $\sigma_0$). The concentrations of separated substances (samples) are assumed to be small enough.
We should emphasize that the coefficients $\alpha_k$ can have different signs (positive or negative); from a physical
viewpoint $\alpha_k<0$ means that this particular substance has lower specific conductivity than that of a buffer
solution. When this substance enters into a solution, it `replaces' the buffer substances and the conductivity of a
mixture is decreasing (for details see
\cite{BabZhukYudE,BelloZhRChrom95,MosherSavilleThorman,ZhEREph96}).

\begin{figure}[H]
  \centering
  \includegraphics[scale=0.8]{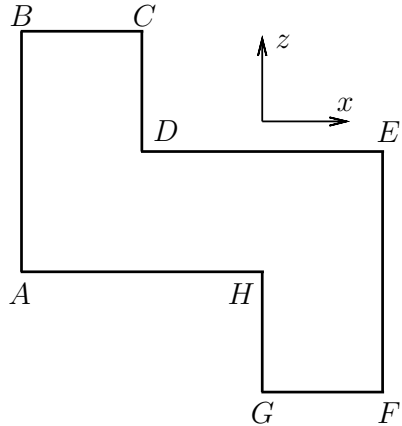}\,
  \caption{Electrophoretic chamber}
  \label{shir_fig1.1}
\end{figure}

Let us consider the domain shown in Fig.~\ref{shir_fig1.1}. We accept that its boundary is rigid with non-leak
conditions for a liquid and for the concentrations $c_k$
\begin{eqnarray}
&& \boldsymbol{v}=0, \quad \boldsymbol{i}_k\cdot
\boldsymbol{n} =0, \quad k=1,\dots,n.
  \label{shir:1.5}
\end{eqnarray}
where $\boldsymbol{n}$ is the unit vector to the boundary.
 The potential $\varphi$ is prescribed on the parts $BC$ and $GF$, while the rest of the boundary is
insulated:
\begin{eqnarray}
&& \varphi\,\big|_{BC}=\varphi_0,  \quad
\varphi\,\big|_{GF}=\varphi_1,
  \label{shir:1.6}
\end{eqnarray}
\begin{eqnarray}
&& \left.\frac{\partial\varphi}{\partial n}\right|_{CDEF,BAHG}=0.
  \label{shir:1.7}
\end{eqnarray}
At the initial instant a fluid is still and the initial distributions of admixtures are prescribed:
\begin{eqnarray}
&& \boldsymbol{v}\big|_{t=0}=0, \quad  c_k\big|_{t=0}=c_k^0(x,z), \quad
k=1,\dots,n.
  \label{shir:1.8}
\end{eqnarray}

\setcounter{figure}{0}

\section{Qualitative Analysis of the Problem}\label{s:2}

\begin{figure}[H]
  \centering
  \includegraphics[scale=0.8]{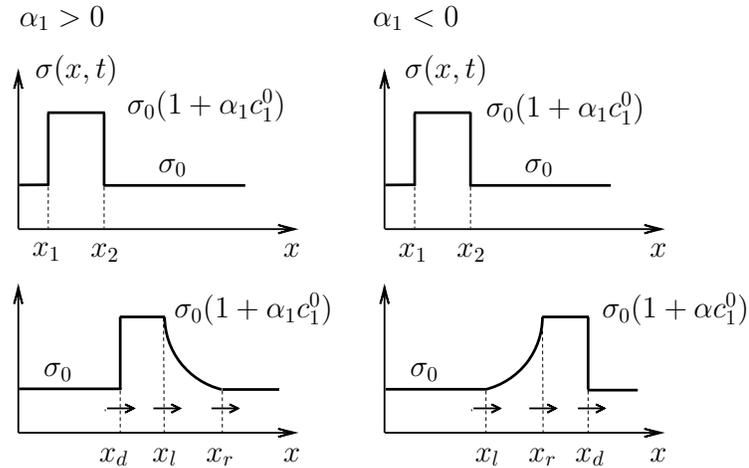}\,
  \caption{The zone evolution}  \label{shir_fig2.1}
\end{figure}

\begin{figure}[H]
  \centering
  \includegraphics[scale=0.8]{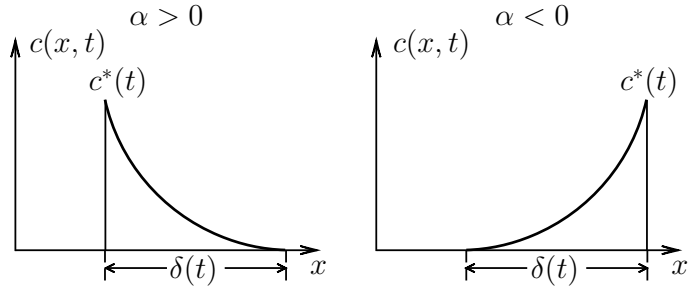}\
  \caption{Final stage of the process}\label{shir_fig2.2}
\end{figure}

There are at least four factors that influence the distortion of a zone as well as admixture concentrations. The first
one is diffusion that causes the spreading of electrophoretic zones. For small concentrations (the case of analytical
electrophoresis) nonlinear effects are weak and diffusion makes the separation of admixtures difficult; its influence
is well studied and described in almost all handbooks on zonal electrophoresis (see
\cite{MosherSavilleThorman,BabZhukYudE}).
The second factor is the electromigration spreading of zones. It reveals itself for high concentrations (for example
for preparative electrophoresis) and described in details in
\cite{BabZhukYudE,BelloZhRChrom95,Zhukov2005}. Recall, that in one-dimensional case the
evolution of an initially rectangular concentration profile for a single admixture follows the patterns shown in
Fig.~\ref{shir_fig2.1} where two upper pictures give initial distributions for positive and negative $\alpha_1$, while
two bottom pictures present some later stages of their developments.
The analysis of diffusionless ($\varepsilon=0$) quasi-linear conservative hyperbolic laws
(\ref{shir:1.2})--(\ref{shir:1.4}) in one-dimensional case with $\boldsymbol{v}=0$ shows that for $\alpha_k<0$ a shock wave
$x=x_d(t)$  at the forward part of concentration profile $c_k$ is formed. Simultaneously there are two fronts of
rarefaction wave $x=x_l(t)$ and $x=x_r(t)$ that appear at the backward part of the profile. The velocities of these
shock wave and the fronts of rarefaction wave are constants:
\begin{eqnarray}
&& V_d=\frac{\gamma_1}{1+\alpha_1 c_1^0}, \quad
   V_l=\gamma_1, \quad
   V_r=\frac{\gamma_1}{(1+\alpha_1 c_1^0)^2}, \quad \alpha_1<0.
  \label{shir:2.1}
\end{eqnarray}
The conductivity in the rarefaction wave is:
\begin{eqnarray}
&& \sigma=\sigma_0\left(
\displaystyle
\frac{\gamma_1 t}{x-x_1}\right)^{1/2}, \quad x_l(t) \leqslant x \leqslant  x_r(t).
  \label{shir:2.2}
\end{eqnarray}
In contrast, for the case $\alpha_k >0$ there is a shock wave $x=x_d(t)$ at the backward part of the profile and there
are two fronts of rarefaction wave $x=x_l(t)$ and $x=x_r(t)$ on the forward part of the profile. The distributions
shown in Fig.~\ref{shir_fig2.1} do exist until the instant $t=t_\textrm{int}$, when a direct interaction between the
waves takes place. For example, for $\alpha_k<0$ the front of rarefaction wave $x_r(t)$ will catch up with the shock
wave $x_d(t)$ \emph{i.e.} $x_r(t_\textrm{int})=x_d(t_\textrm{int})$. The further evolution can be described
analytically. Omitting details, one can notice that finally the concentration profile adapts a `triangular' shape
(Fig.~\ref{shir_fig2.2}).
At $t \to +\infty$ the hight $c^*(t)$ and the base $\delta(t)$ of the `triangle' are given (independently of the sign of $\alpha$) as
\begin{equation}\label{z11.28}
 c^*(t) \sim \left(\frac{M}{\alpha_1 \gamma_1 t}\right)^{1/2}, \quad
  \delta(t)\sim 2\left(\alpha_1\gamma_1 Mt\right)^{1/2},\quad \frac12 c^*(t)\delta(t) \sim M=c_1^0(x_2-x_1),
\end{equation}
where $M$ is the total mass of an admixture. Notice that the `spreading' of the concentration in the absence of
diffusion process is:
\[
c^*(t)=\mathcal{O}(t^{-1/2}), \quad \delta(t)=\mathcal{O}(t^{1/2}), \quad t \to +\infty
\]
It represents a nonlinear effect of electromigration spreading that takes place due to the dependence of the admixture
transfer velocity on the concentration. Notice that in the case of conventional diffusion $c^*(t)$ and $\delta(t)$ are
changing similarly; that is the reason why experimentalists often mix up these two very different phenomena (diffusion
and hyperbolic spreading of a concentration profile); at the same time one should take into account that the spreading
due to diffusion is much slower than hyperbolic spreading (constants in the similar laws are very different!). More
details on the zone evolution the zonal electrophoresis are given in
\cite{BabZhukYudE,Zhukov2005,BelloZhRChrom95,ZhERSIAM}.
The two-dimensional version of this problem can be solved only numerically but we can still analytically derive that
there is a shock wave of concentration on the forward part of the profile and a rarefaction wave of concentration at
the rear part of a wave for $\alpha<0$. A normal coordinate  to a zone boundary corresponds to the $x$-direction in
one-dimensional case.

The third factor causing the distortions of a zone reveals itself only for the domains with vertex points where the
singularities of electric potential take place. In the simplest case when admixtures are absent (pure buffer solution)
and a fluid is still ($\boldsymbol{v}=0$, $c_k=0$, $\sigma=\sigma_0$), the equation $\operatorname{div} \boldsymbol{j}=0$ produces Laplace's equation for
the potential $\Delta\varphi=0$ with the boundary conditions (\ref{shir:1.6}), (\ref{shir:1.7}). In the vicinity of the
vertex of the angle $\alpha$ there is a solution $\varphi=Ar^{\pi/\alpha}\cos(\pi\theta/\alpha)$ where $(r,\theta)$ are
polar coordinates with the origin at a vertex. One can see that the gradient of such a potential possesses a
singularity $\nabla\varphi=\mathcal{O}(r^{\pi/\alpha-1})$. In particular, for the domain shown in
Fig.~\ref{shir_fig1.1} the radial component of electric field $E_r$ at the points $D$ and $H$ ($\alpha=3\pi/2$) is very
large: $E_r=\mathcal{O}(r^{-1/3})$. In contrary, at points $A$ and $E$ ($\alpha=\pi/2$) the singularities are absent
and the electric field is small: $E_r=\mathcal{O}(r)$. Now, let all admixture at the initial instant be placed in the
vicinity of $BC$ and the potential difference $\varphi_1-\varphi_0$ be such that the migration of admixture is directed
towards $GF$. It is apparent that there is a fast admixture transport in the vicinities of points $D$ and $H$, and a
slow one in the vicinities of $A$ and $E$. In the latter case the formation of stagnation regions is likely. This
effect will eventually cause the strong distortion of zone shapes.

Finally, the fourth factor is  gravitational concentration  convection. The difference between the densities of an
admixture and a buffer fluid generates buoyancy flows in the regions of inhomogeneous density. One can guess that there
should be an intense fluid flow induced by the motion of an admixture in the vicinities of points $D$ and $H$ (see
Fig.~\ref{shir_fig1.1}).

\setcounter{figure}{0}

\section{Numerical Experiments}

The problem (\ref{shir:1.1})--(\ref{shir:1.7}) has been solved by a direct simulation with the employing of the
finite-difference method of markers and cells (MAC). In order to approximate the transport equations (\ref{shir:1.2})
we have used combined explicit and implicit finite-difference schemes with the finite differences taken in the
direction opposite to a flow; the latter allows us to block mesh diffusion effects. To compute the flow velocity $\boldsymbol{v}$
we have used explicit schemes. Finally the method of `sequence over relaxation' (SOR) with the relaxation parameter
$1.37\div 1.96$ has been chosen for the computation of pressure $p$ and potential $\varphi$ (in the solving of
finite-difference analogues of elliptic equations). It should be noticed that due to singularities in $\nabla\varphi$
in the vicinities of points $D$ and $H$ the SOR method has been essentially modified: in particular we introduced five
subdomains with the appropriate matching conditions at their boundaries. In addition, the computations have also been
carried out by the finite element method with the use of FreeFem++ and FlexPDE. The comparative relative error between
the computations by the different methods used has been below $0.03$.

The initial concentration is prescribed in a circular zone of radius $r$ centered at $x_0$, $z_0$ as
\[
c^0(x,z)=\frac12\Bigl(1+\tanh\bigl(-\delta( (x-x_0)^2 + (z-z_0)^2-r^2)\bigr)\Bigr),
\]
or in a rectangular zone given by an appropriate expression. Both these distributions represent `almost step functions'
with the parameter $\delta=100$ used for the smoothing of discontinuities.

\subsection{Zone Distortion in Complex Shape Channel}

\begin{figure}[H]
  \centering\includegraphics[scale=1]{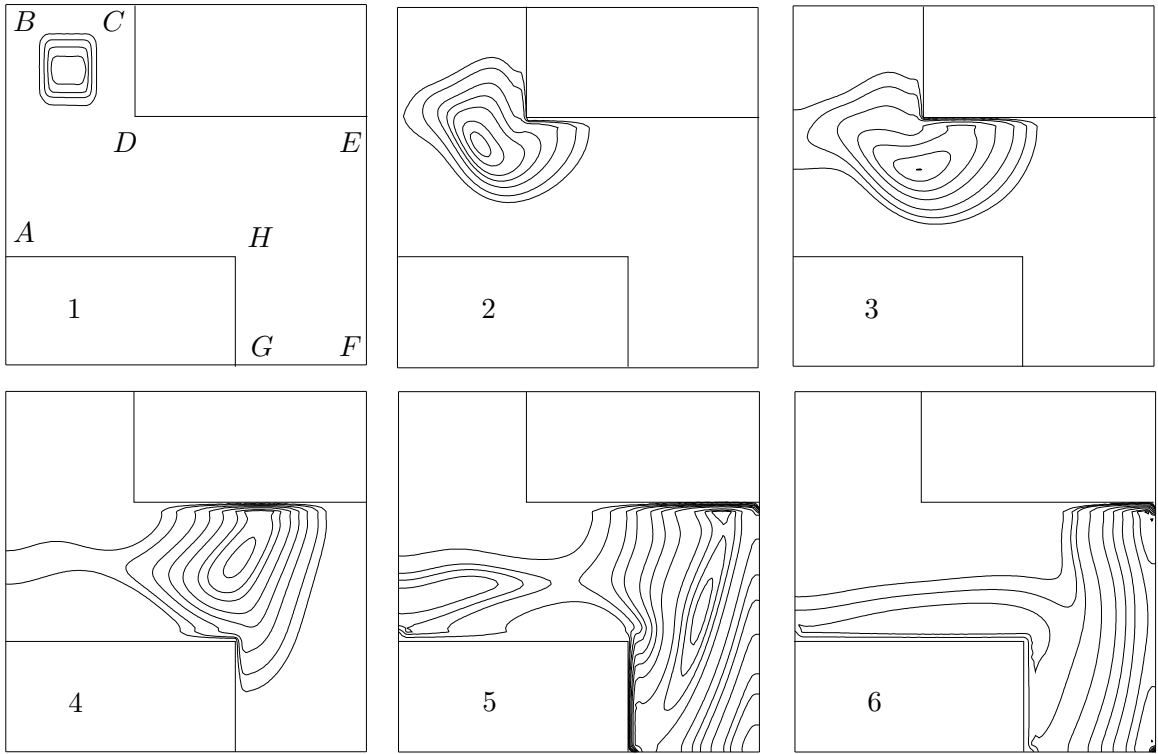}
  \caption{The surface levels of concentration $c_1(x,z,t)$}
  \label{shir_fig3.1}
\end{figure}

\begin{figure}[H]
 \centering\includegraphics[scale=1]{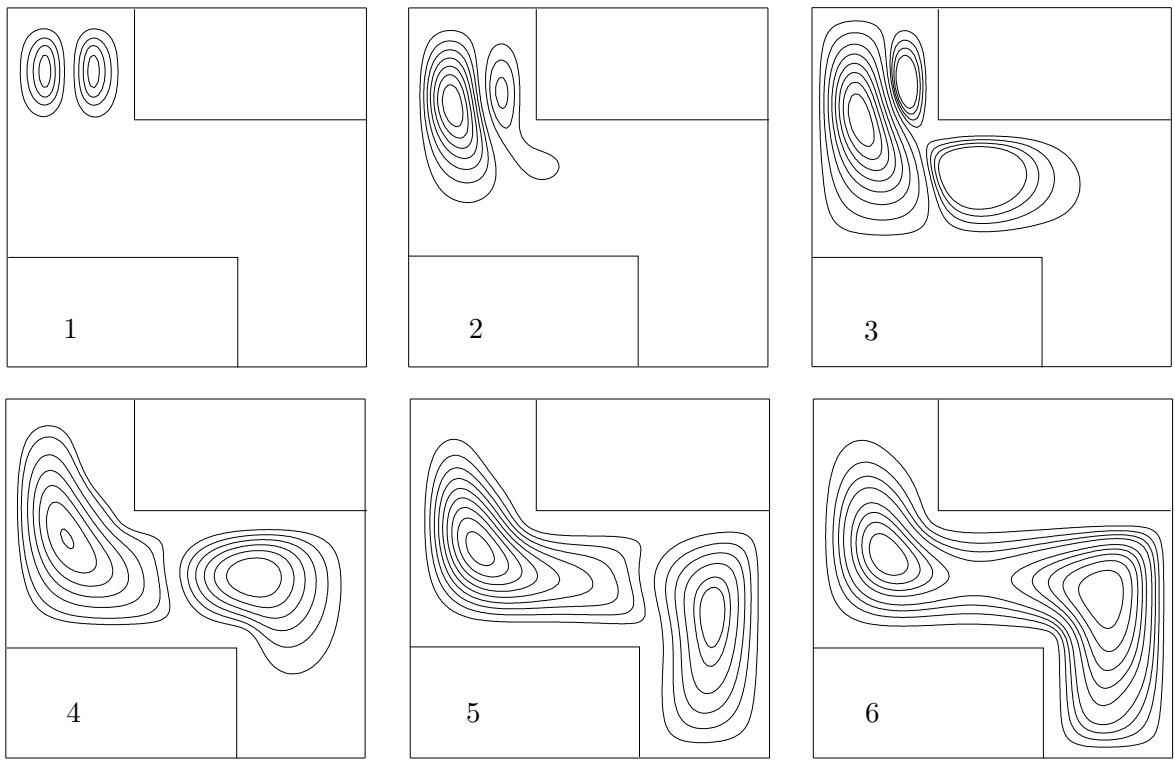}
  \caption{The streamfunction}
  \label{shir_fig3.2}
\end{figure}

\begin{figure}[H]
  \centering
\includegraphics[scale=0.5]{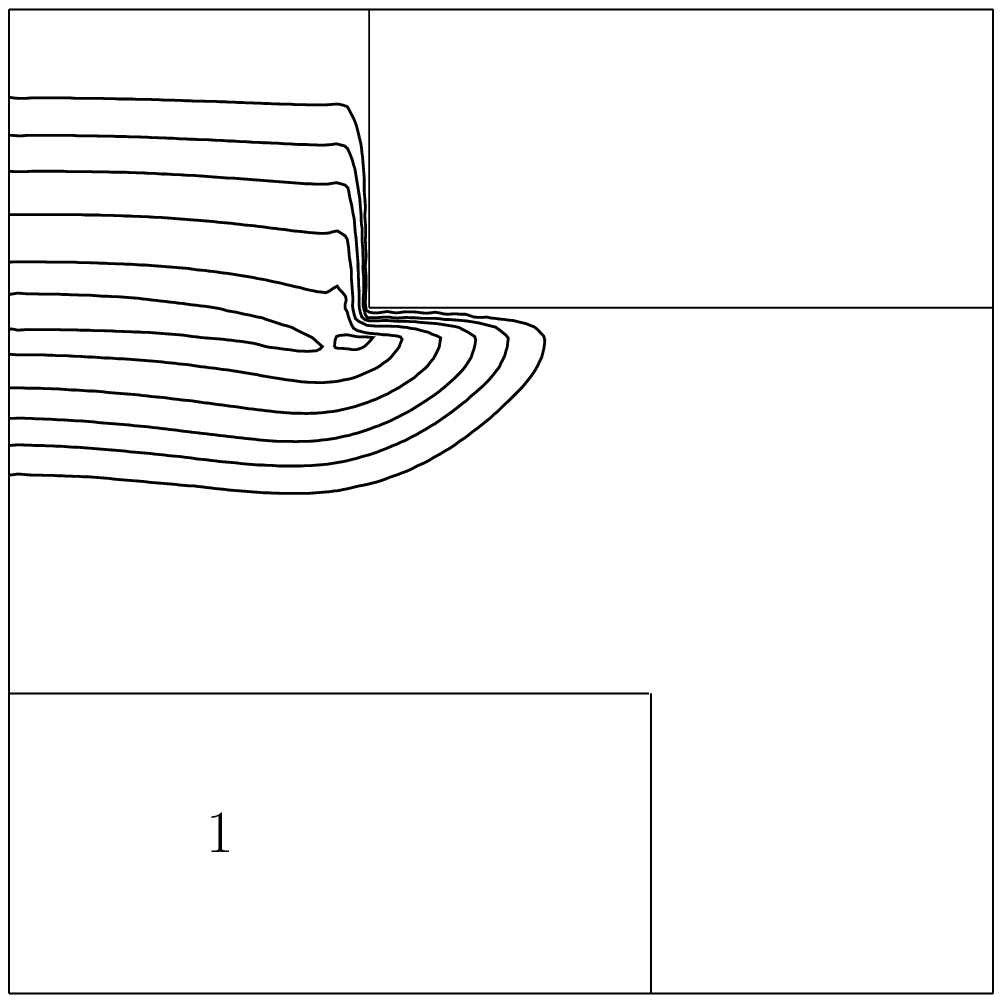}
\includegraphics[scale=0.5]{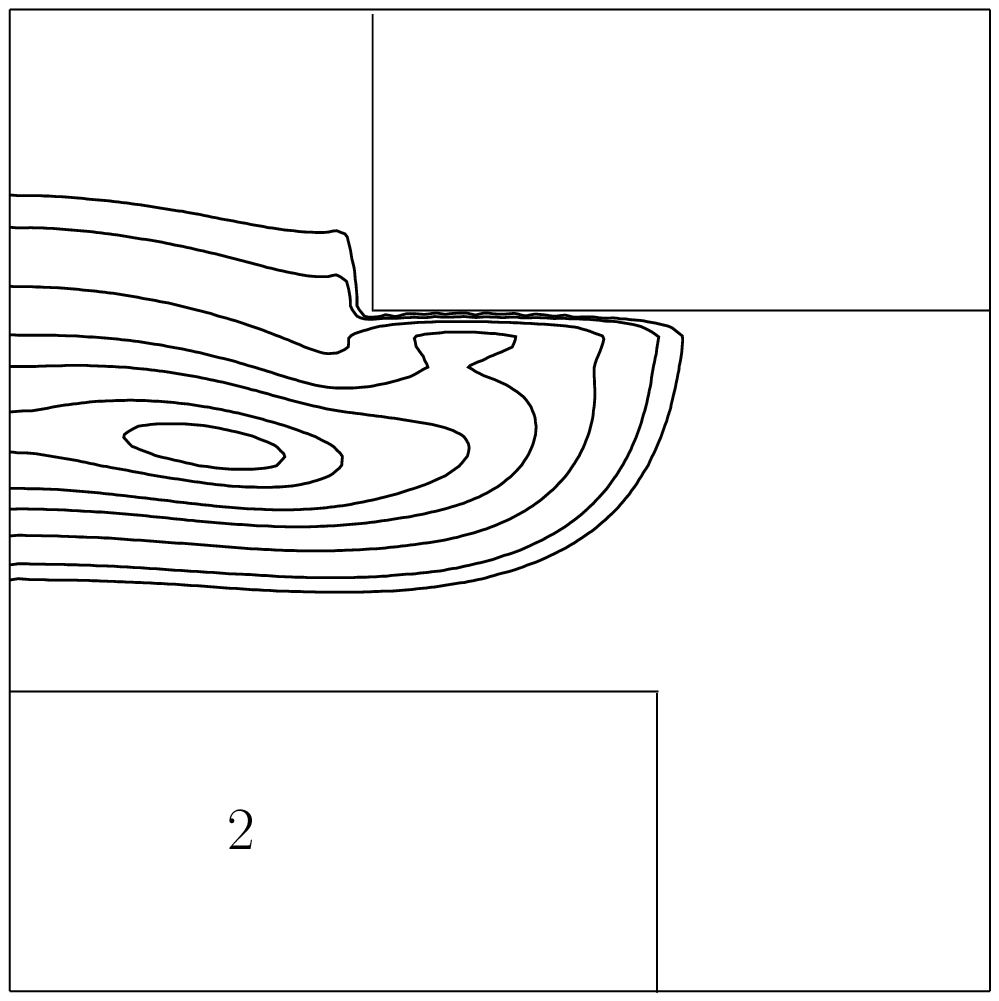}
\includegraphics[scale=0.5]{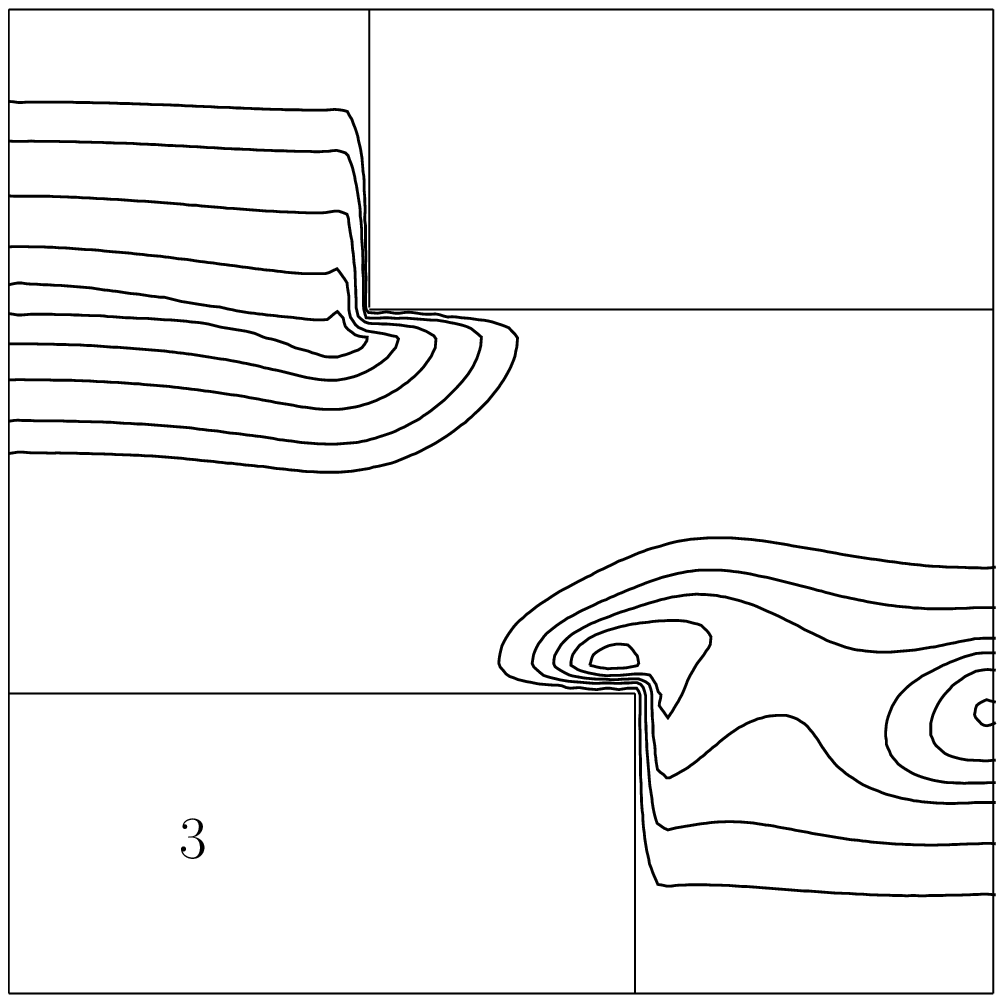}
  \caption{The surface levels of the concentrations $c_1(x,z)$ (frame 1), $c_2(x,z)$ (frame 2), and $c_1(x,z)$, $c_2(x,z)$ (frame 3)}
  \label{shir_fig3.3}
\end{figure}

Some typical results for the channel shown in Fig.~\ref{shir_fig1.1} are presented in
Fig.~\ref{shir_fig3.1}--\ref{shir_fig3.3}. In the case of a lighter single admixture ($\mu=0.01$, $\beta_1=-0.1$,
$\gamma_1=-0.15$, $\varphi_1-\varphi_0=20$, $\alpha_1=-0.4$, $\varepsilon=0.1$) the surface levels of the concentration
$c_1(x,z)$ and the streamlines of fluid flows are shown in Figs.~\ref{shir_fig3.1}, \ref{shir_fig3.2}. The sequence of
frames 1--6 corresponds to the instants $t=0.24, 9.91, 14.66, 26.71, 42.11, 68.50$. The potential difference chosen
drives the zone from the line $BC$ towards the line $GF$. The lengths of intervals in the channel boundary $ABCDEFGH$
are:
\[
|BC|=w_1>0, \ |DE|=1-w_1>0, \ |GF|=w_2>0, \ |AH|=1-w_2>0, \quad w_1=w_2=\frac13,
\]
\[
|CD|=h_1>0, \ |HG|=h_2>0, \ |AB|=1-h_2>0,  \ |EF|=1-h_1>0, \quad h_1=h_2=\frac13.
\]
Combined explicit and implicit schemes have allowed us to use the mesh $60\times 60$ with rather large steps
$h_x=h_y=0.0167$. It is well visible that there are strong distortions of the zone shape in the vicinities of points
$D$, $H$, while in the vicinities of the points $A$, $E$ the admixture is retarded in stagnation regions. It is
interesting that the distortion is so strong that it causes the formation of three vortices (for the instant $t=14.66$
in the frame 3, Fig.~\ref{shir_fig3.2}) in the vicinity of the point $D$ and the subsequent disappearing of one vortex.

The numerical experiments have shown that the variations of the diffusion parameter $\varepsilon$ within the interval
$(0.0001 <\varepsilon< 0.1)$ does not change the results. It means that electromigration (not diffusion!) is the main
reason for the zone spreading, while diffusion can affect the deforming of zone only in large time intervals. A series
of computations have shown that qualitatively similar pictures with the forming of stagnation regions and the
distortion of a zone shape have been observed in a wide interval of $\varepsilon$. The changing of
$|\gamma_1||\varphi_1-\varphi_0|$ (for the fixed $\mu$, $\beta_1$) leads only to the changing of a time scale for the zone
passing through the channel. The typical size of a zone in the direction of motion (before its qualitative distortion
has appeared) has been defined by the parameter $\alpha_1$. It agrees well with the results for one-dimensional case
(see Figs.~\ref{shir_fig2.1}, \ref{shir_fig2.2}) at least for $\alpha_1> -0.5$. The zone distortion is most sensible to
the change of the parameters $h_1$, $h_2$, $w_1$, $w_2$ that prescribe the relative sizes of an electrophoretic
chamber. Any quantitative description of a zone distortion is rather difficult. For `symmetric' chambers with
$0.05<h_1=h_2<0.1$ and $0.05<w_1=w_2<0.1$ up to 40\% of the total mass of an admixture is trapped in the vicinity of
points $A$ and $E$, while for the case $0.3<h_1=h_2<0.4$ and $0.2<w_1=w_2<0.3$ this figure is up to 25\%. The intensity
of convection is strongly influenced by the viscosity $\mu$ or more precisely by  ${\rm Gr}=|\beta_1\mu^{-2}|$, where ${\rm
Gr}$ is a version of Grashof's number related to concentration. An intense (almost chaotic) mixing takes place for
${\rm Gr}> 10^8$.

The frames 1 and 2, Fig.~\ref{shir_fig3.3} show a stage of the separation of two admixtures that are heavier than the
buffer ($\mu=0.01$, $\varepsilon=0.1$, $\varphi_1-\varphi_0=20$, $\beta_1=0.1$, $\beta_2=0.1$, $\gamma_1=-0.15$,
$\gamma_2=-0.35$, $\alpha_1=-0.4$, $\alpha_2=-0.4$). It is interesting to see that the `faster' admixture
($|\gamma_2|>|\gamma_1|$) has been undergoing the larger distortions. We have split the pictures for two concentrations into
two frames in order to avoid visual superimposing of surface levels. The frame 3, Fig.~\ref{shir_fig3.3} shows a stage
of separation of two heavier admixtures moving in the opposite directions (towards each other) ($\mu=0.01$,
$\varepsilon=0.1$, $\varphi_1-\varphi_0=20$, $\beta_1=0.1$, $\beta_2=0.1$, $\gamma_1=-0.15$, $\gamma_2=0.15$,
$\alpha_1=-0.4$, $\alpha_2=-0.4$). It is noticeable that the motion of the admixture $c_2$ against the gravity causes
the greater profile distortions than the motion of the admixture $c_1(x,z)$ along the gravity field.

The described numerical experiments have shown that the motion of zones in the gravity field and in channels with
vertex points does produce very strong zone distortions. This fact explains the failures of the experiments with the
`Kashtan' devices in space that were noticed in \cite{ZhBSazStoyanov,BZhMyshkis}.

\subsection{Zone Distortion in Rectangular Channel}

In order to demonstrate more clearly the influence of convective mixing on the zone distortion we have also made
computations for a simple rectangular channel where the singularities of an electric field are absent. Constant
potentials has been prescribed on side walls, while top and bottom walls has been insulated. The width of the channel
is $a=2$, while its hight is $b=1$. The rest of parameters have been chosen as: $\varphi_1-\varphi_0=20$, $\mu=0.01$,
$\gamma_1=-0.35$, $\varepsilon=0.01$. Below we present the results for a single admixture with different values $\alpha_1$
and $\beta_1$ which show that the gravity can strongly affect the admixture transport via the buoyancy driven intense
vortex flows near the zone.

\begin{figure}[H]
\includegraphics[scale=0.36]{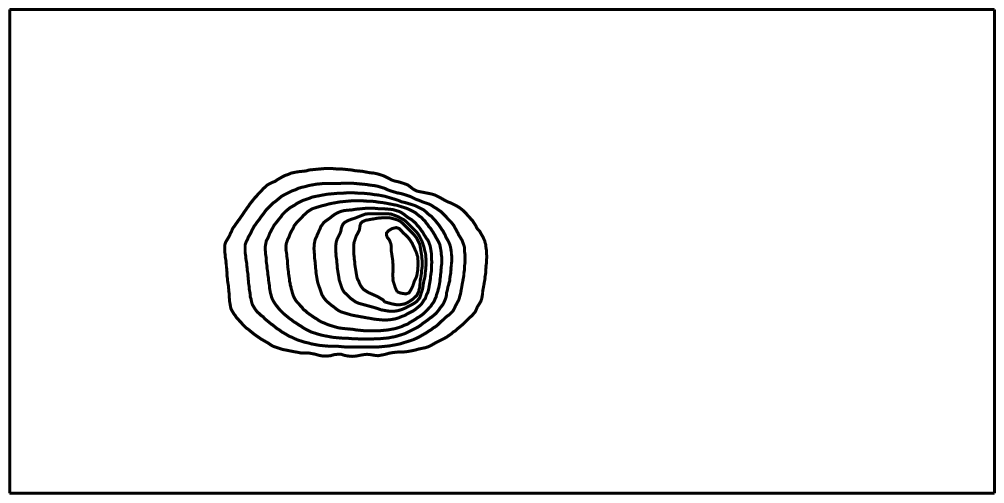}
\includegraphics[scale=0.36]{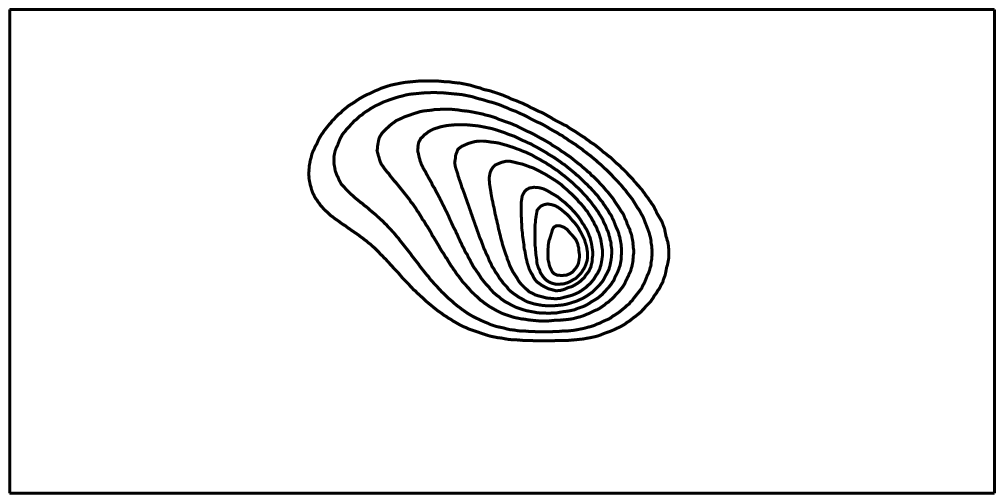}
\includegraphics[scale=0.36]{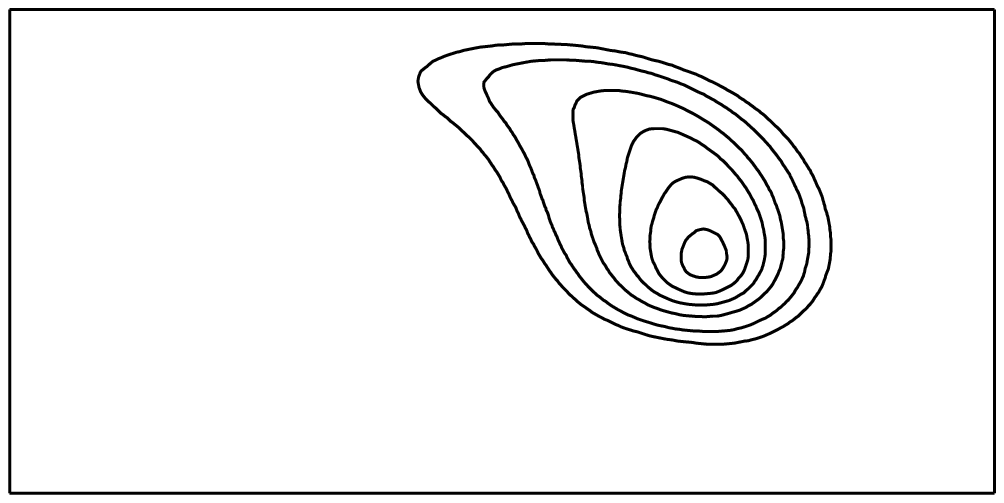}
\includegraphics[scale=0.36]{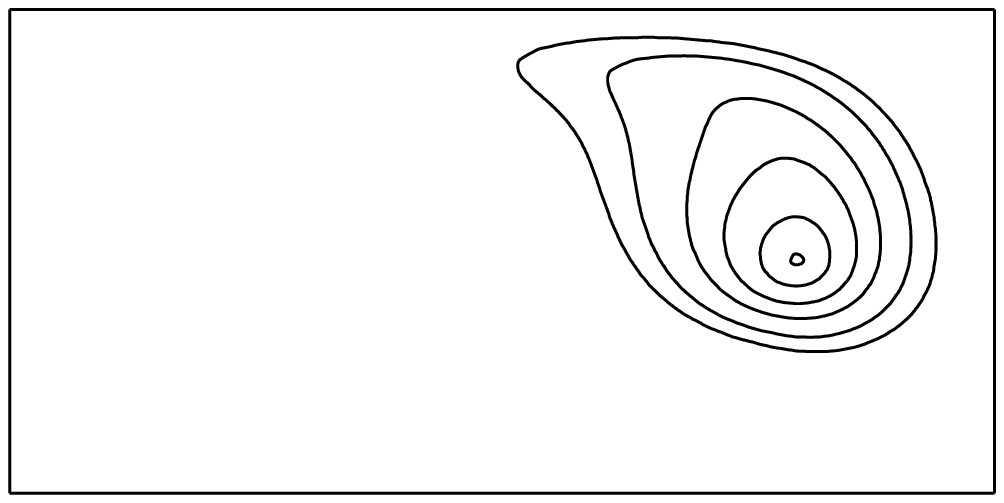}
\caption{The surface levels. $\alpha_1=-0.4$}\label{shir_fig3.4}
\end{figure}

\begin{figure}[H]
\includegraphics[scale=0.36]{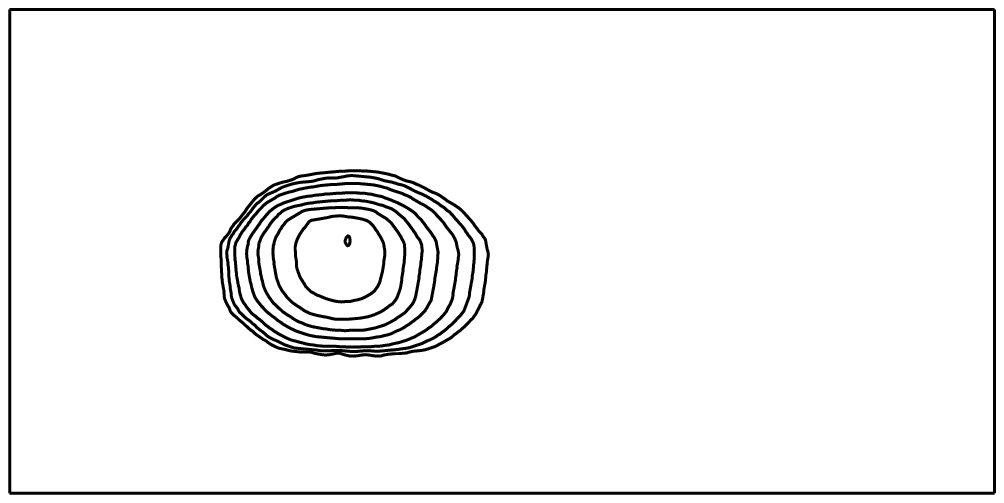}
\includegraphics[scale=0.36]{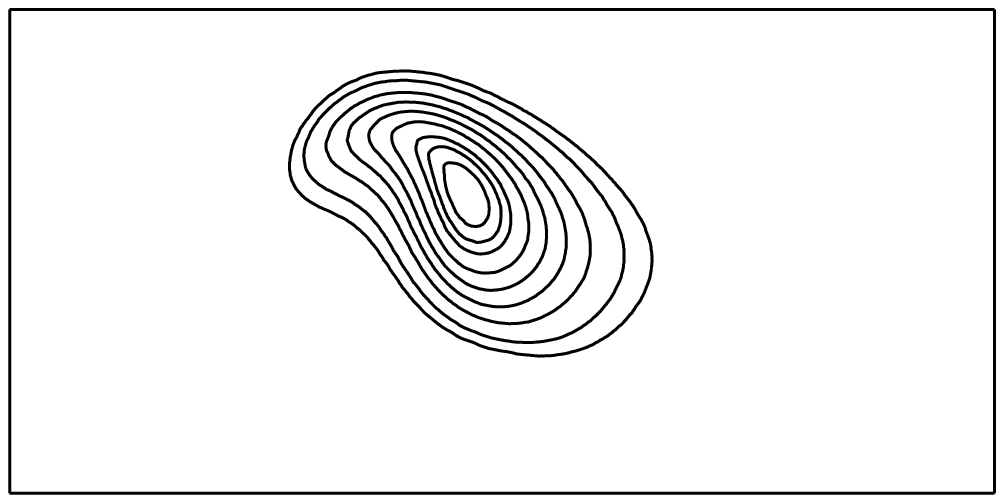}
\includegraphics[scale=0.36]{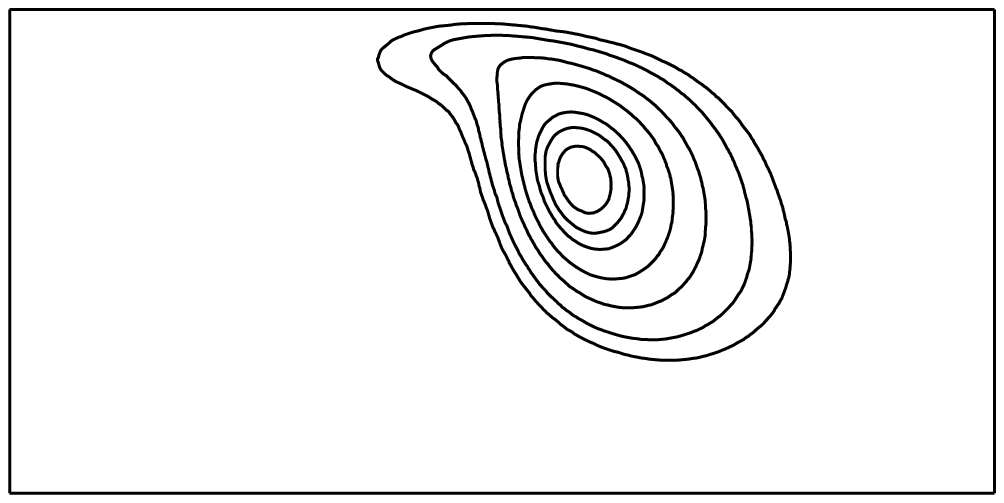}
\includegraphics[scale=0.36]{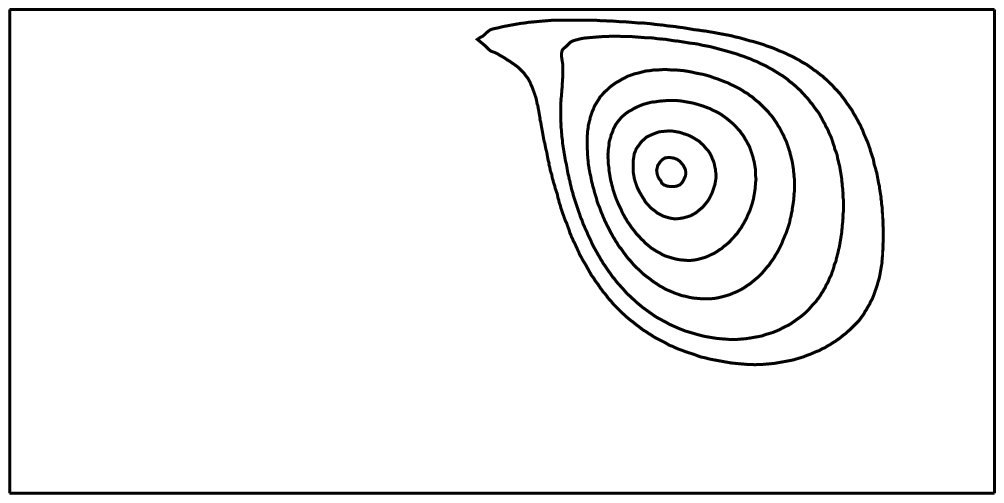}
\caption{The surface levels. $\alpha_1=0.4$}\label{shir_fig3.5}
\end{figure}

Figs.\,\ref{shir_fig3.4}, \ref{shir_fig3.5} present the surface levels for $\beta_1=-200$ (a lighter admixture) in a
fluid with small viscosity (${\rm Gr}=2\cdot10^{6}$) for two different values of $\alpha_1$  at the instants
$t=0.0256$; $0.1024$; $0.1792$; $0.2304$ where one can clearly see electromigration spreading. For $\alpha_1<0$
(Fig.\,\ref{shir_fig3.4}) the forward front exhibits the compression of surface levels (a shock wave), while the rear
front~--- a rarefaction wave. For $\alpha_1>0$ (Fig.\,\ref{shir_fig3.5}) those features are opposite: the rear front
represents a shock wave, while the forward front~--- a rarefaction wave. These results have been in agreement with the
properties of motion in one-dimensional case (see Figs.\,\ref{shir_fig2.1}, \ref{shir_fig2.2}). Notice that the zone
form has been essentially distorted at the initial stages of motion. The buoyancy effects at $\beta_1=-200$ are very
weak; the admixture is slightly moving upwards while it is transported by an electric field.

\begin{figure}[H]
\centering
\includegraphics[scale=0.48]{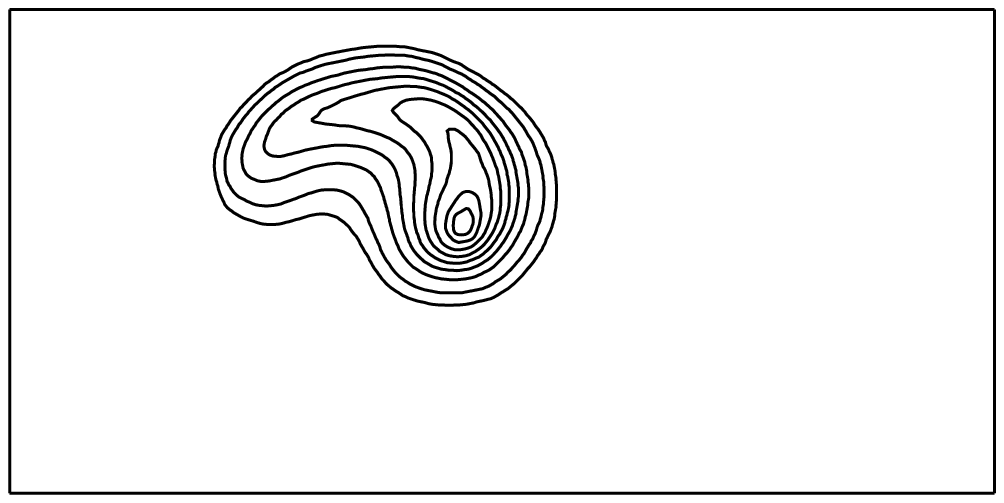}
\includegraphics[scale=0.48]{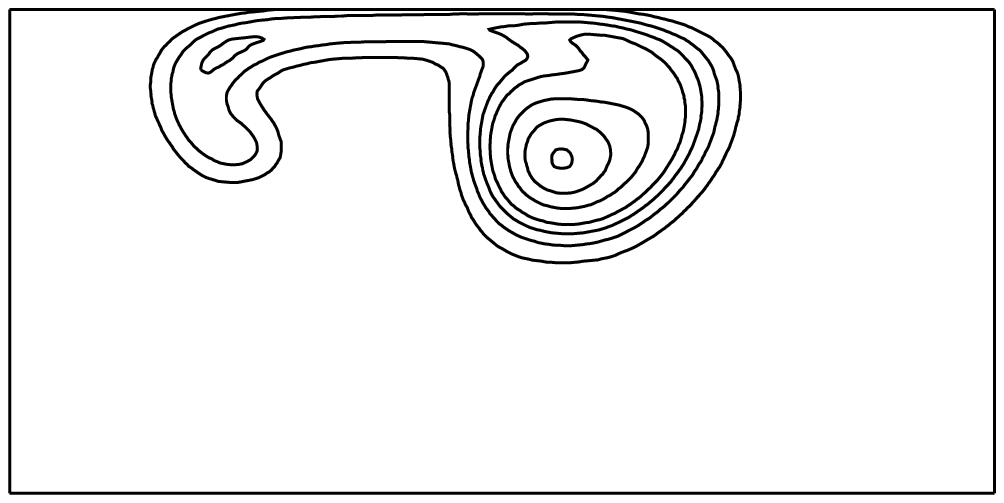}
\includegraphics[scale=0.48]{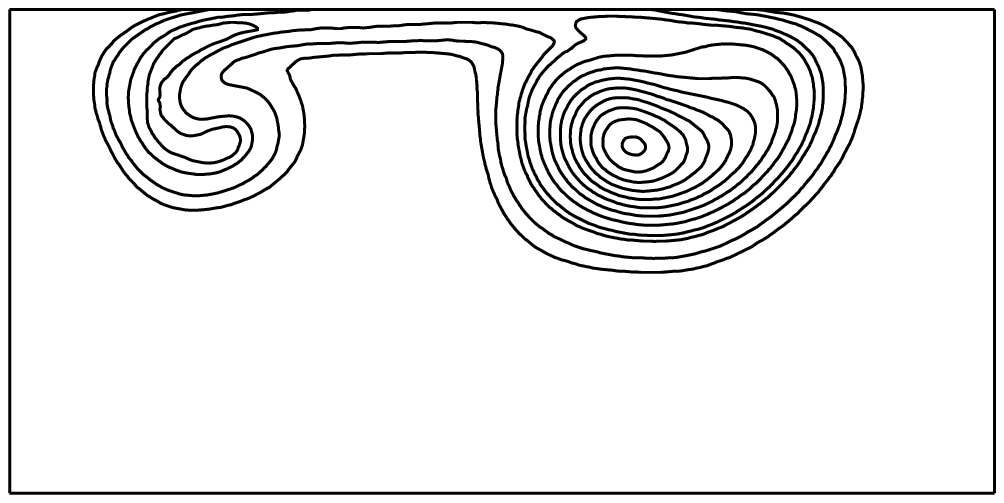}
\caption{The surface level of concentration}\label{shir_fig3.6}
\end{figure}

Fig.\,\ref{shir_fig3.6} corresponds to $\alpha_1=-0.4$. It shows the surface levels of concentrations at the instants
$t=0.0512$; $0.1024$; $0.1280$. Here we have chosen the admixture being five times lighter ($\beta_1=-1000$, ${\rm
Gr}=2\cdot10^{7}$) than in the previous case of Figs.\,\ref{shir_fig3.4}, \ref{shir_fig3.5}; so the influence of the
gravity here is quite essential. One can clearly see the strong interaction of the arising zone with the upper wall
that causes the apparent splitting of a single zone into two (Fig.\,\ref{shir_fig3.6}). The next
Fig.\,\ref{shir_fig3.7} demonstrates the correspondent streamline pictures that reveal the forming of a vortex pair in
the process of convective mixing. These results show that the gravity can change the zone shape drastically.
\begin{figure}[H]
\centering
\includegraphics[scale=0.48]{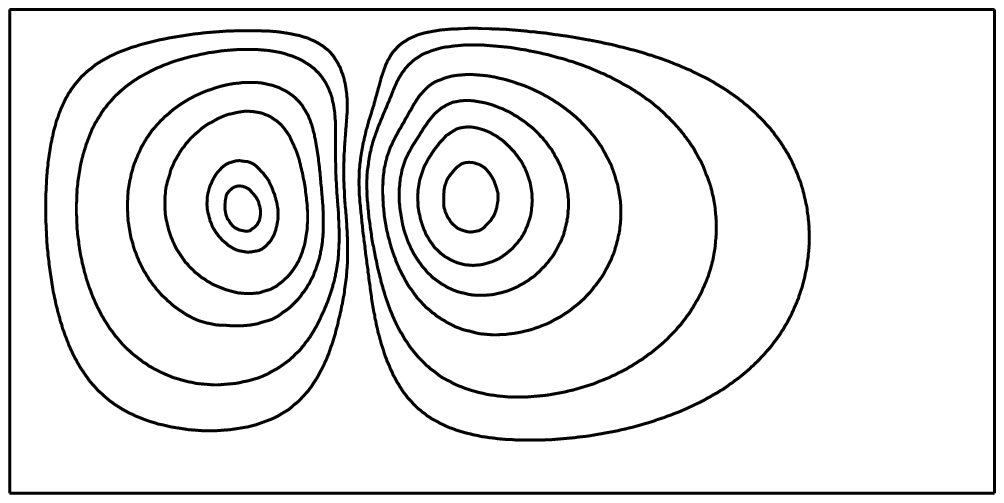}
\includegraphics[scale=0.48]{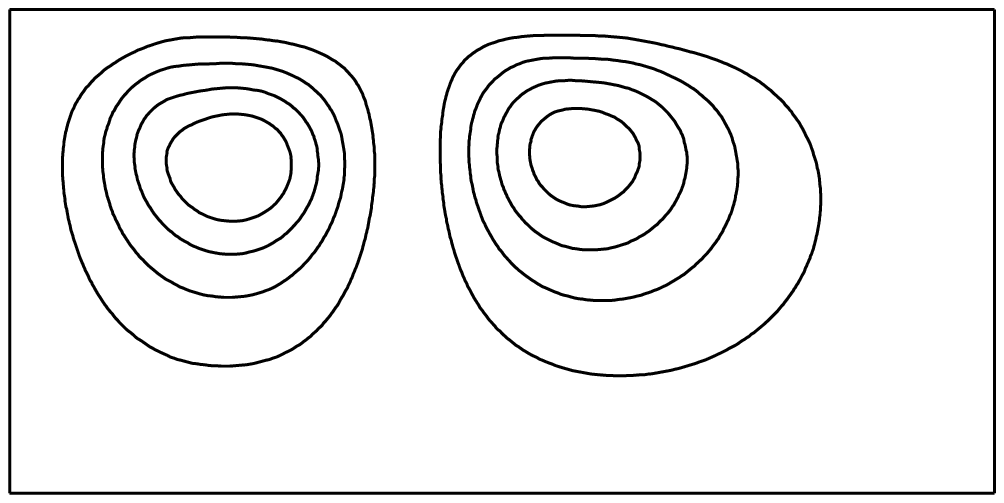}
\includegraphics[scale=0.48]{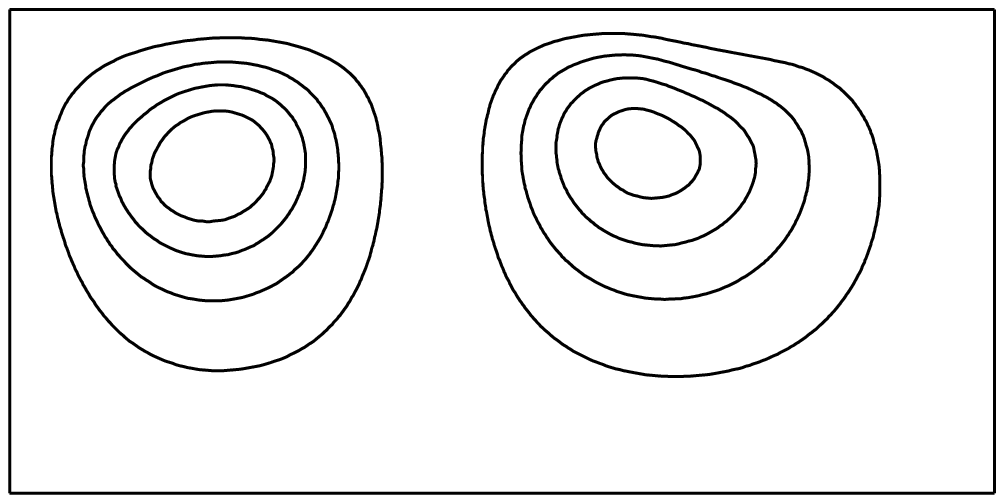}
\caption{The streamlines}\label{shir_fig3.7}
\end{figure}

\section*{Acknowledgments}

This research is partially supported by EPSRC (research grants
GR/S96616/01, EP/D055261/1, and EP/D035635/1), by the Russian
Ministry of Education (programme `Development of the research
potential of the high school',  grants 2.1.1/6095 and 2.1.1/554),
and by Russian Foundation for Basic Research (grants 07-01-00389,
08-01-00895, and 07-01-92213 NCNIL). The authors are grateful to the
Department of Mathematics of the University of York for the
providing of excellent conditions for this research.

\newpage
\renewcommand{\bibname}{\centering{\large\bf References}}


\begin{thebibliography}{100}
\bibitem{KanianskyMasarBodor}
Kaniansky,~D.; Masar,~M.; Bodor,~R.; Zuborova,~M.;  Olvecka,~E.; Johnck,~M.; Stanislawski,~B.
\emph{Electrophoresis.} \textbf{2003}, \emph{24} (12--13), 2208--2227.

\bibitem{EricksonLiuKrullLi}
Erickson,~D.; Liu,~X.; Krull,~U.\,J.; Li,~D.
\emph{Anal. Chem.} \textbf{2004}, \emph{76}, 7269--7277.

\bibitem{BharadwajSantiagoMohammadi}
Bharadwaj,~R.; Santiago,~J.\,G.; Mohammadi,B.
\emph{Electrophoresis.} \textbf{2002}, \emph{23}, 2729--2744.

\bibitem{MolhoHerrMosier}
Molho,~J.\,I.; Herr,~A.\,E.; Mosier,~B.\,P.; Santiago,~J.\,G.; Kenny,~Th.\,W.
\emph{Anal. Chem.} \textbf{2001}, \emph{73}, 1350--1360.

\bibitem{JenWuLinWu}
Jen,~C.; Wu,~C.; Lin,~Y.; Wu,~C.
\emph{Lab Chip.} \textbf{2003}, \emph{3}, 77--81.

\bibitem{JohnsonRossLocascio}
Johnson,~T.\,J.; Ross,~D.; Locascio,~L.\,E.
\emph{Anal. Chem.} \textbf{2002},  \emph{74}, 45--51.

\bibitem{BerliPiaggioDeiber} 
Berli,~C.\,L.\,A.; Piaggio,~M.\,V.; Deiber,~J.\,A.
\emph{Electrophoresis.} \textbf{2003}, \emph{24} (10), 1587--1595.

\bibitem{ChenSantiago1} 
Chen,~C.-H.; Santiago,~J.\,G.
In Proc. \emph{IMECE};  2002, \emph{1} (33563).

\bibitem{EricksonLi1} 
Erickson,~D.; Li,~D.
\emph{Langmuir.} \textbf{2003}, \emph{19}, 5421--5430.

\bibitem{ErmakovNano} 
Ermakov,~S.\,V.; Jacobson,~S.\,C.; Ramsey,~J.\,M.
In Tech. Proc. of the Int. Conference on Modeling and Simulation of Microsystems MSM 99;  U.S.A., 1999; pp.\,534--537.

\bibitem{ErmakovNano1} 
Ermakov,~S.\,V.; Jacobson,~S.\,C.; Ramsey,~J.\,M.
\emph{Anal. Chem.} \textbf{1998}, \emph{70}, 4494--4504.

\bibitem{ErmakovJacobsonRamsey2000} 
Ermakov,~S.\,V.; Jacobson,~S.\,C.; Ramsey,~J.\,M.
\emph{Anal. Chem.} \textbf{2000}, \emph{72}, 3512--3517.

\bibitem{Ghosal47} 
Ghosal,~S.
\emph{Electrophoresis.} \textbf{2004}, \emph{25} (2), 214--228.

\bibitem{GuijtEvenhuisMackaHaddad} 
Guijt,~R.\,M.; Evenhuis,~C.\,J.; Macka,~M.; Haddad,~P.\,R.
\emph{Electrophoresis.} \textbf{2004}, \emph{25} (23--24), 4032--4057.

\bibitem{Herr} 
Herr,~A.\,E.; Molho,~J.\,I.; Drouvalakis,~K.\,A.; Mikkelsen,~J.\,C.;  Utz,~P.\,J.; Santiago,~J.\,G.; Kenny,~Th.\,W.
\emph{Anal. Chem.} \textbf{2003}, \emph{75}, 1180--1187.

\bibitem{HuWernerLi} 
Hu,~Y.; Werner,~C.; Li,~D.
\emph{Anal. Chem.} \textbf{2003}, \emph{75}, 5747--5758.

\bibitem{OddyMikkelsenSantiago} 
Oddy,~M.\,H.; Mikkelsen,~J.\,C.; Santiago,~J.\,G.
\emph{Anal. Chem.} \textbf{2001}, \emph{73}, 5822--5832.

\bibitem{PatankarHu} 
Patankar,~N.\,A.; Hu,~H.\,H.
\emph{Anal. Chem.} \textbf{1998}, \emph{70} (9),  1870--1881.

\bibitem{PatankarSantiago} 
Patankar,~N.\,A.; Santiago,~J.\,G.
\emph{Anal. Chem.} \textbf{2005}, \emph{77},  6672--6781.


\bibitem{Santiago4} 
Santiago,~J.\,G.
\emph{Anal. Chem.} \textbf{2001},  \emph{73},  2353--2365.

\bibitem{BabZhukYudE} 
Babskii,~V.\,G.; Zhukov,~M.\,Yu.; Yudovich,~V.\,I.
\emph{Mathematical theory of electrophoresis}; Plenum Publishing Corporation: New York, 1989.

\bibitem{ZhYuDAN} 
Zhukov,~M.\,Yu.; Yudovich,~V.\,I.
\emph{Soviet Physics Doklady.} \textbf{1982}, \emph{27}, 918--924.

\bibitem{ZhJVM84} 
Zhukov,~M.\,Yu.
\emph{J. Comp. math. and math. phys.} \textbf{1984}, \emph{24}
(4), 549--565 (in Russian).

\bibitem{ZhEREph96} 
Ermakov,~S.\,V.; Zhukov,~M.\,Yu.; Righetti,~P.\,G.
\emph{Electrophoresis.} \textbf{1996}, \emph{17}, 1134--1142.

\bibitem{BelloZhRChrom95} 
Bello,~M.\,S.; Zhukov,~M.\,Yu.; Righetti,~P.\,G.
\emph{J. Chrom. A.} \textbf{1995}, \emph{693}, 113--130.

\bibitem{Zhukov2005} 
Zhukov,~M.\,Yu.
\emph{Mass transfer in an electric field}; Rostov University Press: Rostov-on-Don, 2005.

\bibitem{ZhERSIAM} 
Zhukov,~M.\,Yu.; Ermakov,~S.\,V.; Righetti,~P.\,G.
\emph{SIAM J. Appl. Math.} \textbf{1999}, \emph{59} (2), 743--776.

\bibitem{ZhECRAnal94} 
Ermakov,~S.\,V.; Zhukov,~M.\,Yu.; Capelli,~L.; Righetti,~P.\,G. \emph{Anal. Chem.} \textbf{1994}, \emph{66}, 4034--4042.


\bibitem{BZhMyshkis} 
Babskii,~V.\,G.; Zhukov,~M.\,Yu.; Myshkis,~A.\,D.; Kopachevskii,~N.\,D.; Slobozhanin,~L.\,A.; Tyuptsov,~A.\,D.
\emph{Methods of Solving Problems of
Hydromechanics in Zero Gravity}; Naukova Dumka: Kiev, 1992.

\bibitem{ZhBSazStoyanov} 
Zhukov,~M.\,Yu.; Babskii,~V.\,G.; Sazonov,~L.\,I.; Stoyanov,~A.\,V. \emph{Space science and technology.} \textbf{1989}, \emph{4}, 15--19 (in Russian).

\bibitem{Bello} 
Bello,~M.\,S.; Polezhaev,~V.\,I.
\emph{Fluid Dynamics.} \textbf{1990}, 2, 14--20 (in Russian).

\bibitem{Polezaev-3} 
Polezhaev,~V.\,I.; Bello,~M.\,S.; Verezub,~N.\,A.
\emph{Convection in zero gravity.} Nauka: Moscow, 1991 
(in Russian).

\bibitem{ZhZiva94} 
Zhukov,~M.\,Yu.; Tsyvenkova,~O.\,A.
\emph{Fluid Dynamics.} \textbf{1994}, \emph{29} (5), 717--723.

\bibitem{ZhZiva95} 
Zhukov,~M.\,Yu.; Tsyvenkova,~O.\,A.
\emph{Fluid Dynamics.} \textbf{1995}, \emph{30} (5), 652--660.

\bibitem{ZhPetr97MZhG} 
Zhukov,~M.\,Yu.; Petrovskaya,~N.\,V.
\emph{Fluid Dynamics.} \textbf{1997}, \emph{32} (5), 631--641.

\bibitem{ZhSazonovDifUr97} 
Zhukov,~M.\,Yu.; Sazonov,~L.\,I.
\emph{Diff. Uravn.} \textbf{1997}, \emph{3} (4), 470--477 (in Russian).

\bibitem{MosherSavilleThorman} 
Mosher,~R.\,A.; Saville,~D.\,A.; Thorman,~W.
\emph{The dynamics of electrophoresis}; VCH Publishers: New York, 1992. \end{thebibliography}
\end{document}